\documentclass[10pt,leqno]{amsart}
\usepackage[utf8]{inputenc}
\usepackage{graphicx}
\usepackage{indentfirst,csquotes}
\usepackage{amssymb}
\usepackage{etoolbox}
\usepackage{hyperref}

\hypersetup{colorlinks=true,linkcolor=black,filecolor=black,urlcolor=black}

\makeatletter
\renewcommand{\@setaddresses}{%
  \par\nobreak\begingroup
  \centering\normalfont\footnotesize
  \interlinepenalty\@M
  \def\author##1{}%
  \def\address##1##2{\noindent##2\par}%
  \def\email##1##2{%
    \noindent\textit{Email address:}\space
    \href{mailto:##2}{\nolinkurl{##2}}\par}%
  \addresses
  \endgroup
}
\apptocmd{\@setauthors}{\@setaddresses}{}{%
  \PackageError{manuscript}{Could not add affiliations to the title block}{}%
}
\let\enddoc@text\@empty
\makeatother

\begin{document}
\title{Localized Joule heating and carbonization process optimization of 3D printed polymer structures for lab-on-CMOS applications} 
\author[Haque, Nizam, Lavrik, and McFarlane]{Mohammad Aminul Haque{$^1$}, Andalib Nizam{$^1$}, Nickolay Lavrik{$^2$}, and Nicole McFarlane{$^1$}}
\date{\today}
\address{{$^1$} Electrical Engineering \& Computer Science, University of Tennessee, Knoxville, TN, USA.\\
{$^2$} Center for Nanophase Materials Sciences, Oak Ridge National Lab, Oak Ridge, TN, USA}
\email{anizam@alum.utk.edu}

\begin{abstract}
Use of carbon electrodes integrated with CMOS technology has great promise for lab-on-chip, biomedical, environmental, and industrial applications. One method of developing carbon electrodes is to 3D print polymer nanostructures on a CMOS chip, and use a thermal annealing process to pyrolyze them into carbon at high temperature, approximately 950 $^\circ$C. Such high temperature is detrimental to the CMOS sensor circuits, causing severe degradation. We present a novel solution of localized Joule heating of these polymer nanostructures to achieve complete carbonization of the electrodes. Both analytical and COMSOL simulations guided the experiments to characterize the carbonization process conditions. Partial carbonization was achieved at 500--600 $^\circ$C of global temperature, and complete carbonization was achieved by localized Joule heating. Pyrolized nanostructures monolithic samples on a silica wafer were fabricated using standard material processing, such as 2-Photon Polymerization. Pyrolysis was achieved by Rapid Thermal Processing (RTP). Observations using optical microscopy, scanning electron microscopy, and Raman spectroscopy demonstrated the feasibility of polymer carbonization within a CMOS compatible temperature window. 
\end{abstract} 

\maketitle

\section{Introduction}
\begin{figure}[t]
    \centering
    {\includegraphics[width=3in]{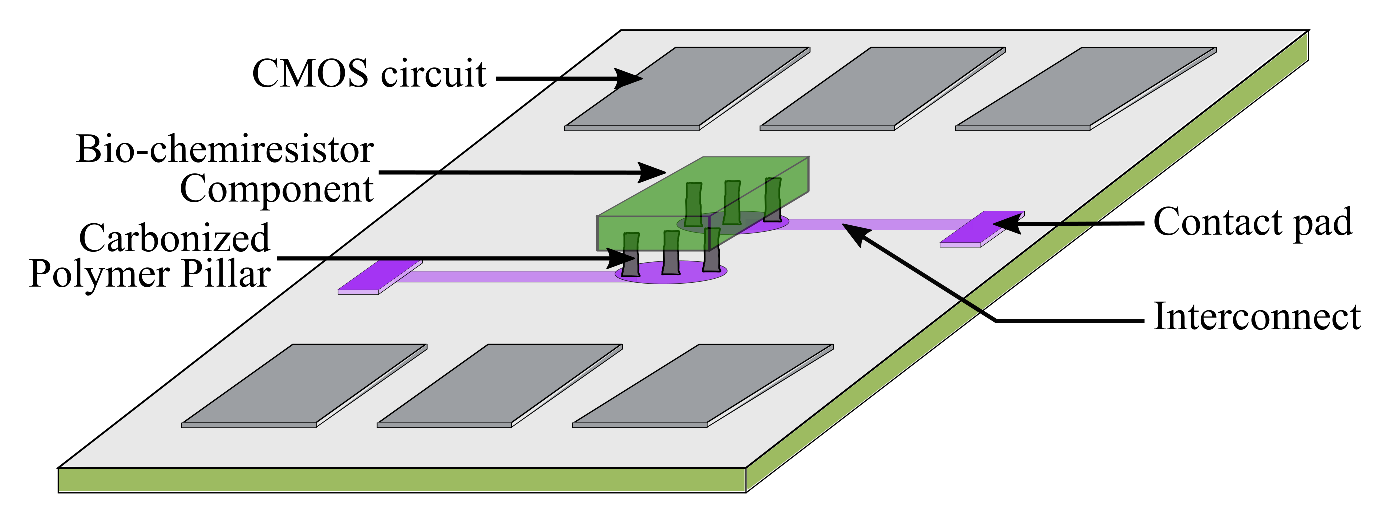}
    \caption{A tentative lab-on-CMOS system depicting the position of the isolated live biological cells placed in contact with the carbonized polymer nanoelectrodes. The electrodes are connected to contact pads on both sides. The readout CMOS circuit contains the designed analog-front end.}
    \label{fig00}}
\end{figure}

Top-down fabrication methods such as 3D printing of polymer precursors followed by pyrolysis is a recently developed approach to develop biocompatible carbonized microelectrodes \cite{sun20232d,shao2022carbon,liu2023biocompatible,luong2020gram}. However, processing conditions, such as the high temperatures, used for pyrolysis hinder the integration of carbonized microelectrodes with widely-used complementary metal oxide semiconductor (CMOS) technologies. This limits the development of lab-on-CMOS biosensors. An illustration of a lab-on-CMOS biosensor system is shown in figure \ref{fig00}. As such, this work focused on the optimization of the process parameters, leading to CMOS-compatible fabrication temperature to achieve complete carbonization of polymer precursors through localized Joule heating.

Different forms of carbon-based structures have historically been used for electrochemical sensing. Carbon nanotubes and graphene have been used to detect ammonia, bisphenol A, dopamine, ascorbic acid, and other biochemical molecules. Carbon nanofibers have been used to detect glucose, caffeic acid, dopamine, H$_2$S, and serotonin \cite{rani2020nimoo4,sakthivel2018entrapment,saunier2020carbon,zhang2018zno}. The established vertically aligned carbon nanofiber (VACNF) include catalytic chemical vapor deposition (C-CVD) and plasma enhanced chemical vapor deposition (C-PECVD) \cite{melechko2005vertically,al2016glucose}. The synthesis of carbon nanotubes uses arc discharge, laser ablation, and CVD. Graphene can be produced from both graphitic and nongraphitic sources using chemical-mechanical exfoliation, epitaxial growth, and CVD \cite{novoselov2004electric,hernandez2008high,emtsev2009towards,li2009large,terasawa2012growth,whitener2014graphene}. On-chip integration of carbon-based electrodes with CMOS electronics has been a focus of recent research due to the ability to create compact, low power, intelligent miniaturized sensing systems \cite{dudina2019carbon,dudina2019monolithic,kim2021stress,SONG2022139570}. 3D printing-based solutions have also become attractive for integration. A high-frequency commercial CMOS IC has been designed using a 3D printed metal pipe rectangular waveguide. A mechanically flexible CMOS temperature sensor with 3D printed carbon nanotube has been designed. In the biosensor field, soft and implantable sensors, microfluidic biosensors, and wearable haptic sensors have been designed using various 3D printing techniques \cite{zhu20233,honda2015high,ali2022recent}. However, the fabrication of these carbon-based electrodes has not been done using pyrolysis. Such direct fabrication of carbon-based electrodes through pyrolysis in the vicinity of a CMOS sensor circuit requires more than 950 $^\circ$C. Such high temperatures will damage the CMOS circuit's functionality. We have previously developed protocols for the carbonization of 3D printed polymer precursors and have found that pyrolysis of up to 500  $^\circ$C provides partial pyrolysis, while complete carbonization requires temperatures of 900  $^\circ$C \cite{haque2020carbonization,haque2019carbonized,haque2019carbonized2,haque2021carbonized3}. These high temperatures are detrimental to the on-chip CMOS circuitry \cite{ranganathan2000photoresist,schueller1997fabrication,wang2005novel}. Although localized Joule heating itself is not a novel process, it has not been used in the purpose of carbonization of polymer structures. Our proposed approach is as follows. A two-step annealing process achieved partial carbonization, which was followed by localized Joule heating through those partially carbonized structures. This step achieves complete carbonization of these polymer structures. In this work, this novel approach to implement localized Joule heating in partially carbonized polymers to achieve complete carbonization.

In this paper, the focus is on current-driven localized heating (also known as Joule heating) of the partially carbonized conducting polymer structure to convert it to a completely carbonized polymer. While these localized heating methods have been shown to be very promising for a variety of breakthrough applications \cite{mamleyev2019laser,zang2020laser,mori2014electrically}, to the best of our knowledge, they have not been applied to carbonization of 3D printed polymer structures. Localized heating will allow protection of the CMOS circuitry. 

Semiconducting and switching behaviors of partially pyrolyzed polymers have been historically observed. These materials can show small signal resistivity of 1 to $10^6$ $\Omega$-cm as well as a temperature-dependent change in resistivity. However, the use of semiconducting properties of partially pyrolyzed precursors to obtain complete pyrolysis has not been explored. In this work, a fabrication procedure for partial pyrolysis followed by localized Joule heating is presented. This paper expands upon prior work \cite{haque2020carbonization,haque2019carbonized,haque2019carbonized2,haque2021carbonized3}. The novelty of this work lies in exploring the irreversible change in conductivity resulting from localized Joule heating in a 3D printed partially carbonized structure.
The paper is organized as follows. In Section II, the experimental fabrication process is outlined.  Section III describes the analytical approximations used, and Section IV outlines the results. Finally, we provide concluding remarks in Section V.

\section{Experiment}
\begin{figure}[t]
    \centering
    {\includegraphics[width=3in]{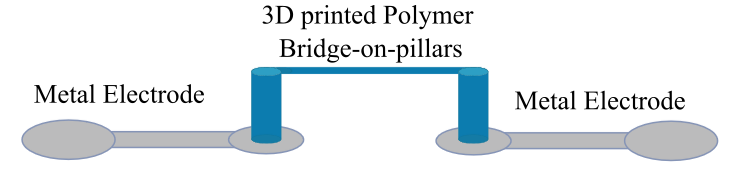}
    \caption{Illustration of 3D-printed bridge-on-pillars (BoP) structure on metal electrodes.}
    \label{fig000}}
\end{figure}

In order to facilitate significant Joule heating within a partially carbonized polymer structure, 3D printed bridge-on-pillars (BoP) structures on an electrode–patterned silica substrate were fabricated \cite{haque2021carbonized3}. An illustration of this structure is shown in figure \ref{fig000}. The structure consists of two pillars connected by a thin bridge. Metal leads connect from the pads to the pillars for electrical connectivity. A platinum-on-titanium stack was patterned using contact photolithography and metal lift off on a 100 mm silica wafer. The sample chips were obtained by dicing the wafer using an Accretech SS10 dicing saw (15 mm $\times$ 15 mm sample size). Titanium and platinum were used to provide adhesion and thermal stability, respectively.

BoP structures were 3D printed on metalized silica substrates using a commercially available two-photon polymerization Nanoscribe tool (Photonic Professional GT, Nanoscribe GmbH) \cite{carlotti2019functional}. Three sets of BoP structures of various aspect ratios were fabricated using the vendor supplied IPS photopolymer. Structure A-2530183 consists of a Pillar that is 2 $\mu$m in height and 4 $\mu$m in diameter, with a Bridge that is 30 $\mu$m in length and 0.183 $\mu$m in diameter. It is noteworthy here that cavitation (bubble formation) can occur if the laser power is not optimized. 3D laser writing was followed by developing the 3D structures using SU-8 developer for 15–20 minutes, bathing the chip in isopropanol, and blow-drying with filtered nitrogen.

\begin{figure}[t]
    \centering
    \includegraphics[width=3in]{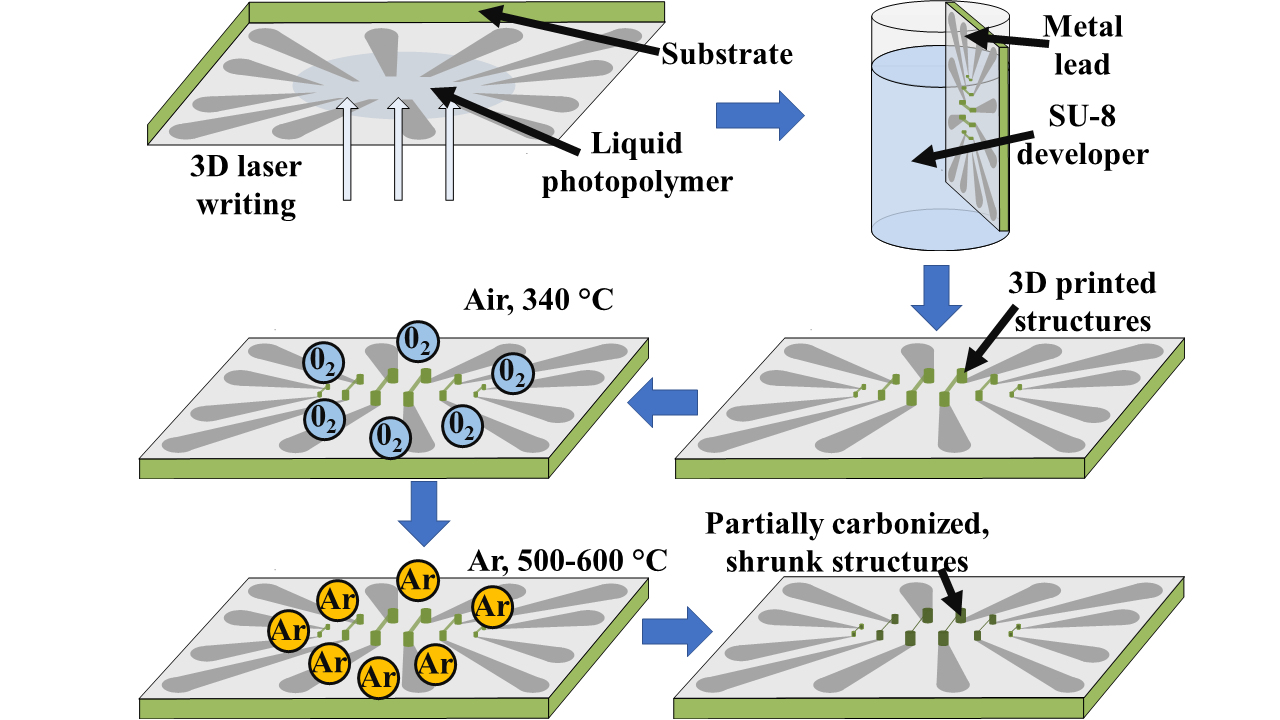}
    \caption{Process sequence to obtain partially pyrolyzed electrically conducting BoP structures on silica substrates. 10 nm titanium (Ti)/50 nm platinum (Pt) stack was used as metal leads to facilitate electrical biasing. Structures were initially 3D printed using two-photon polymerization method and then developed. A CMOS-compatible, two-step annealing process was followed to obtain partially carbonized BoP structures {\cite{haque2021carbonized3}.}}
    \label{fig1}
\end{figure}

The pyrolysis process is consistent with our previous work on polymer carbonization, and includes a two-step annealing process. The first step includes heating the chip and holding it at 340 $^\circ$C for 10 minutes in presence of air on a heated stage (Linkam Scientific LTS420E-P) to initialize oxidation. The second step contains the carbonizing anneal in argon (Ar), or hydrogen ($H_2$) and Ar combined at 500-950 $^\circ$C for 10 minutes to 27 hours. This process sequence is shown in figure \ref{fig1}. Before and after various processing steps, a scanning electron microscope (SEM, Zeiss Merlin) was used to image the structures and Raman spectroscopy (Renishaw InVia) was used to determine the degree of carbonization.

Conductivity of the bridge section of the BoP structure, $\sigma$ was calculated using the following formula,
\begin{equation}
  \sigma=1/\rho  
\end{equation}

\noindent where, $\rho$ is the resistivity and is,
\begin{equation}
\rho=RA/L    
\end{equation}

\noindent where, $R$ is the resistance, $A$ is the cross-sectional area, and $L$ is the length of the bridge. To determine $R$, the voltage was swept from -9 V to +9 V across the BoP structure (in series with a current limiting resistor) through the metal leads using an electrochemical system (PARSTAT 2273, Princeton Applied Research) and the corresponding current was recorded. The cross-sectional area, A was calculated using, 

\begin{equation}
A=\pi r^2
\end{equation}

where, $\pi$=3.14159, and r is the radius of the bridge. L and r were determined from SEM images of the annealed BoP structures.

Conductivity at varying temperatures of the sample was observed by increasing temperature of the chip from room temperature (23 $^\circ$C) to 400 $^\circ$C using a Linkam Scientific LTS420E-P heated stage. To observe the change in conductivity under a constant voltage bias, a fixed voltage of 31.6 V was used with a series 50 k$\Omega$ current limiting series resistance of 50 k$\Omega$ resistor. The current-voltage relationship of the partially pyrolyzed polymer at high temperature was observed using the resistor and a current limiting series resistance of 50 k$\Omega$ and a voltage sweep of 5 V to 31.6 V (upper limit of the voltage source).

\section{Analytical Approximations}

\begin{figure}[t]
    \includegraphics[width=3in]{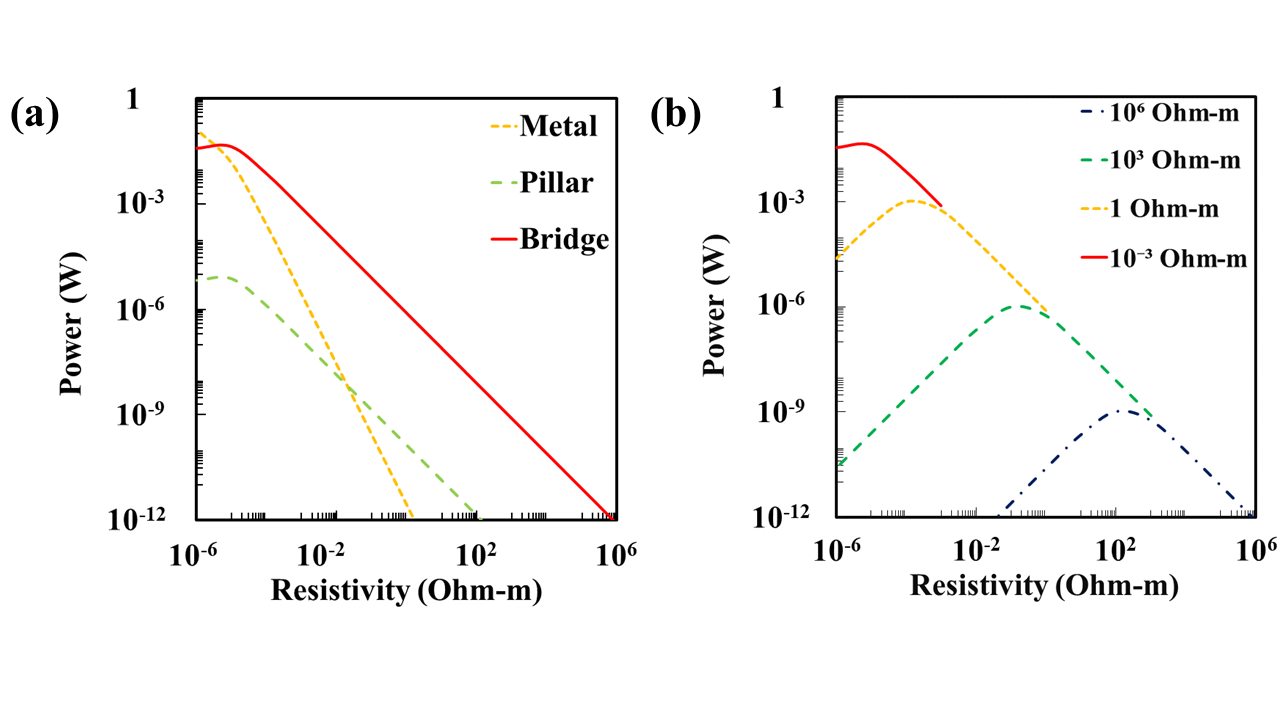}
    \caption{a) Comparison of estimated electrical power on metal leads, pillars, and bridge section as a function of bridge and pillar resistivity of the electrically conducting 3D bridge-on-pillars structure for assumed geometrical dimensions (Pillar). (b) Estimated electrical power at the bridge section of the bridge-on-pillars structure as a function of decreasing bridge resistivity from the same starting bridge and pillar resistivity for assumed geometrical dimensions (A-2530183).}
    \label{fig2}
\end{figure}

Analytical approximations were explored to approximate the electrical power delivery at different sections of the system, namely the bridge, pillars, and metal leads. The objective was to determine the required experimental conditions to obtain complete pyrolysis of partially pyrolyzed polymer (PPP) structure through the Joule heating approach. For this, a thin PPP wire bridge was considered to be between two PPP pillars connected to two metal electrodes. Although the starting precursor material of both the bridge and pillars were the same, the geometrical parameters were different. Bridges were of smaller diameter, thereby having larger resistance and producing higher joule heating. On the other hand, pillars were of larger diameter. Metal electrodes were used to facilitate voltage biasing. Using experiment-based realistic geometrical parameters, a passive component (resistor) based model of the experimental structure was developed. Within a safe voltage limit of 30 V and for a wide range of resistivity values, this stationary model was used to estimate current flow and electrical power at metal, pillars and bridge sections of the system. The following assumptions were considered:
\begin{itemize}
    
\item This is a steady-state system.
\item	Unidirectional heat flow is considered.
\item	No internal heat is being generated within the pillars.
\item	Thermal conductivity is a constant, and it does not vary with temperature. Based on the literature, we considered a thermal conductivity value of 2 W/mK \cite{zhu2018carbonization,venkatachalam2017thermal}.
\item	Contact resistances (electrical and thermal) were ignored.
\item	No radiative heat changes were considered.
\end{itemize}	

The total current through the structure, I, was calculated using Ohm's Law,
\begin{equation}
  I=\frac{V}{R}  
\end{equation}
where, $V$ is the applied voltage bias, and $R$ is the total resistance calculated using given by,
\begin{equation}
    R=R_{metal}+R_{pillars}+R_{bridge}
\end{equation}
where, $R_{metal}$, $R_{pillars}$, and $R_{bridge}$ were the resistances of metal, pillars, and bridge, respectively. $R_{metal}$, $R_{pillars}$, and $R_{bridge}$ were determined by,
\begin{equation}
   R_{metal}=(\rho_{metal} L_{metal})/(H_{metal} W_{metal} ) 
   \end{equation}
   \begin{equation}
  R_{pillars}=2\times(\rho_{pillar} L_{pillar})/(\pi r_{pillar}^2 )      
   \end{equation}
\begin{equation}
R_{bridge}=(\rho_{bridge} 
L_{bridge})/(\pi r_{bridge}^2 )
\end{equation}
where, $\rho_{metal}$, $\rho_{pillar}$, and $\rho_{bridge}$ were the resistivities of metal, pillar, and bridge, respectively, $L_{metal}$, $L_{pillar}$, and $L_{bridge}$ are the length of metal, pillar, and bridge, respectively, $H_{metal}$ and $W_{metal}$ are the thickness and width, respectively, of the metal. $r_{pillar}$ and $r_{bridge}$ are the radii of pillar and bridge, respectively.

The electrical powers of metal ($P_{metal}$), pillars ($P_{pillar}$), and bridge ($P_{bridge}$) were,
\begin{equation}
  P_{metal}=I^2 R_{metal}
  \end{equation}
  \begin{equation}
P_{pillar}=I^2 R_{pillars}
\end{equation}
\begin{equation}
P_{bridge}=I^2 R_{bridge}  
\end{equation}

\begin{figure}[t]
    \centering
    \includegraphics[width=3in]{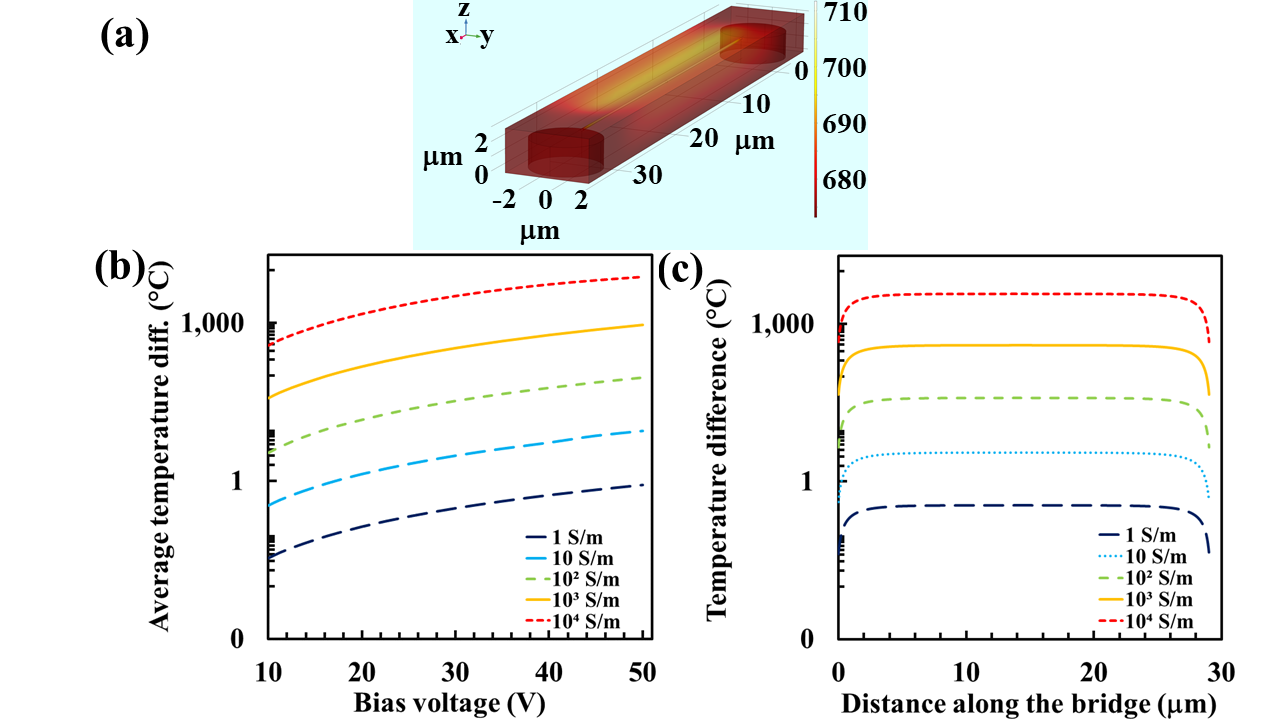}
    \caption{(a) Heat map of the COMSOL model of bridge-on-pillars structure having the dimensions of A-2530183, electrical conductivity of 100 S/m, and under a bias voltage of 30 V. (b) Simulated average difference between temperatures within the bridge section of the bridge-on-pillars structure and ambient 400 $^\circ$C as a function of applied bias voltage across the pillars. Each curve corresponds to a starting electrical conductivity of both bridge and pillar. (c) Simulated difference between the local temperature of the bridge and ambient 400 $^\circ$C along the bridge section of the bridge-on-pillars structure for an applied bias voltage of 30 V. Each curve corresponds to a starting electrical conductivity of both bridge and pillar. In all the cases of (a), (b), and (c), thermal conductivity of bridge and surrounding gas were 2 and 0.05 W/mK, respectively, and thermal emissivity of bridge was 0.8 W$m^2$. }
    \label{fig3}
\end{figure}

Finite-element based COMSOL simulation was conducted to determine the temperature gradient within the structure in an environment approximating the experimental condition. For this, a gaseous environment was considered surrounding the structure to replicate experimentally present flowing forming gas (2\% $H_2$ + 98\% Ar). To simulate heat loss from the structure surface through the surrounding gas, radiation and convection (gas velocity 0 $ms^{-1}$) was considered while surface emissivity of the bridge section was considered to be 0.8 W$m^2$. To take into account the change in the thermal conductivity as a function of increasing temperature of the bridge-on-pillars structure due to applied bias, temperature dependent thermal conductivity of the surrounding gas was considered. Heat capacity at constant pressure, relative permittivity, density, and thermal conductivity of the bridge-on-pillars structure were 1420 J/(kg$\cdot$K), 1, 1190 kg/$m^3$, and 2 W/mK. 

\section{Results}
Before current is driven through the bridge-on-pillar structure, it is important to ensure that higher power will be generated within the active bridge section compared to the metal and pillars. An approximation of the generated power within the bridge was also required to estimate the feasibility of the designed approach as they are annealed and resistivity is reduced by driving current from a reasonable, safe voltage.
In order to estimate the power using analytical approximation, a realistic set of geometrical parameters with a  pillar (2 $\mu$m in height, 4 $\mu$m in diameter) and bridge (30 $\mu$m in length, 0.183 $\mu$m in diameter) was assumed for the bridge-on-pillars  structures. These parameters were obtained from estimated shrinkage in polymer precursors after pyrolysis compared to the 3D printed geometry as  observed in previous work \cite{haque2020carbonization}.

The analytical approximation in figure \ref{fig2}(a) shows the power generated in metal leads, pillars, and bridge sections for the same bridge and pillar resistivities of a wide range of $10^{-6}$ to $10^6$ $\Omega$-m and an applied voltage of 30 V. 

Due to the bridge being designed thinner than the pillars section, the bridge generates higher power compared to both metal and pillars, for bridge and pillar resistivities as low as of 10$^{-4}$ $\Omega$-m. Power generation at the metal is the same order of magnitude as the bridge only when their resistivities converge, specifically within two orders of magnitude.

Figure \ref{fig2}(b) shows the power generation at the bridge as a function of the anticipated resistivity change for a starting bridge-pillar resistivity. When the starting resistivity of both the bridge and pillars is $\approx1$ $\Omega$-m for both the bridge and pillar, the reduced resistivity of the bridge can generate up to microwatts of power. This implies that a resistivity of less than 1 $\Omega$-m (or greater than 1 S/m of conductivity) can potentially generate significant power within the active bridge section for complete carbonization using Joule heating by applying a voltage of 30 V. Aside from the estimated power, the generated temperature at the bridge is also important to determine the required operating conditions (such as the length of the bridge, required conductivity of starting material, and required bias voltage).

Figure \ref{fig3}(a) shows a heat map of the simulation model with geometrical dimensions of the sample and under a bias voltage of 30 V.   It shows that higher temperature generation is mostly concentrated near the bridge region. Our simulation shows that in the presence of a surrounding gaseous environment, heat dissipation normal to the bridge surface is dominated by heat dissipation through the gas and not through the pillars as heat sinks. This heat dissipation is affected by the thermal conductivity of the surrounding gas. 

\begin{figure}[t]
    \centering
    \includegraphics[width=3in]{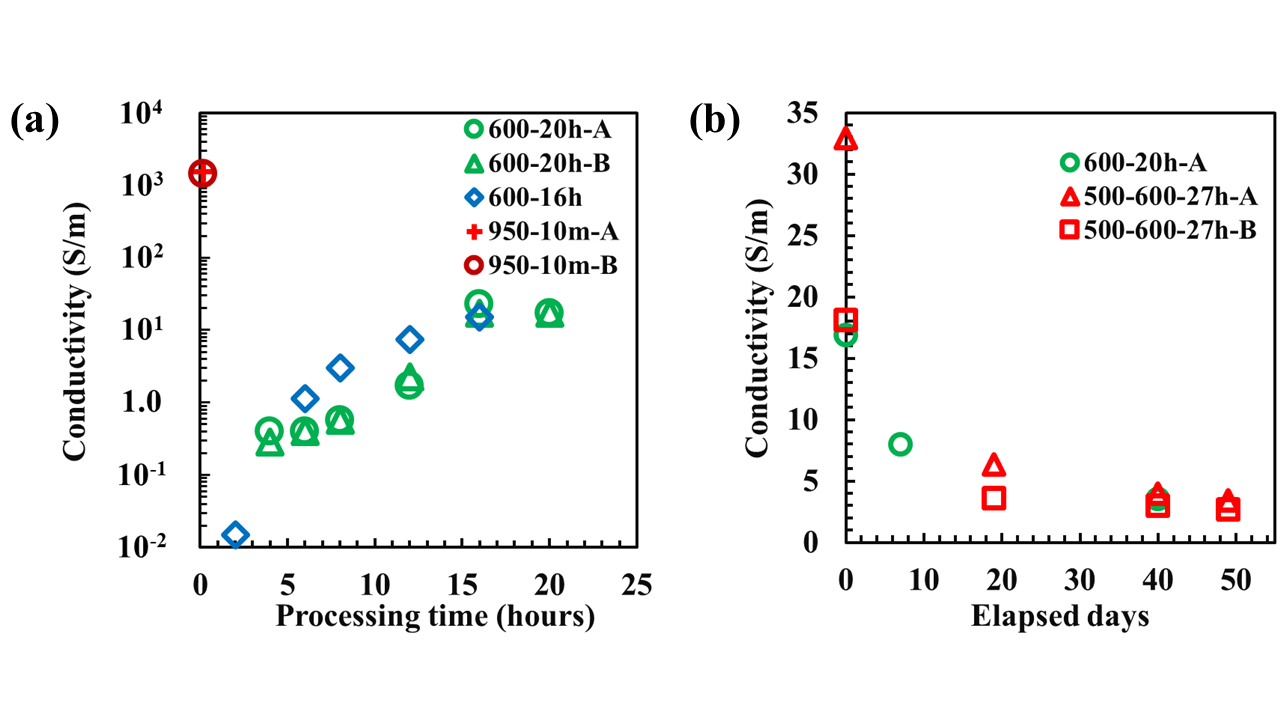}
    \caption{(a) Measured conductivity of the bridge section of bridge-on-pillars structures of samples 600-20h-A, 600-20h-B, 600-16h, 950-10m-A, and 950-10m-B as a function of annealing time at 600 $^\circ$C and 950 $^\circ$C in RTP. (b) Drop in measured bridge conductivity over time of samples 600-20h-A, 500-600-27h-A, and 500-600-27h-A.}
    \label{fig4}
\end{figure}

An estimation of average temperature gradient between the bridge and surrounding gas is useful in understanding the feasibility of the approach. Thus, figure \ref{fig3}(b) shows average temperature difference between bridge and surrounding gas as a function of applied bias voltage. It shows that the average temperature gradient goes above 330 $^\circ$C at an applied voltage of 30 V when electrical conductivity of the bridge and pillar sections are 1000 S/m. While the experimental setup would allow a safe voltage limit of 31.6 V, the simulation shows that increased bias voltage generates higher average temperature gradient.

The temperature gradient along the bridge is helpful in realizing the region of high temperature generation. Figure \ref{fig3}(c) shows temperature difference between the bridge and surroundings along the bridge length. Different curves correspond to increasing electrical conductivity of the structure. It shows that a temperature gradient of more than 400 $^\circ$C is achievable when electrical conductivity is above 1000 S/m.

The bridge-on-pillars structures need to be electrically conductive enough to drive significant current through it in order to generate high temperatures from Joule heating. As such,  bridge-on-pillars structures were annealed in the First Nano Rapid Thermal Processor (RTP) for varying temperature and time, as shown in Table 1.

\begin{table*}
        \renewcommand{\arraystretch}{1.2}
        \caption{Processing sequence to obtain electrically conducting bridge-on-pillar structures.}
        \label{1table}
        \centering
      \setlength\tabcolsep{3pt}
        \footnotesize
        \begin{tabular}{|p{1in}|p{4in}|}
            \hline
            \textbf{Sample} & \textbf{Processing steps} \\
            \hline
            600-20h-A & Anneal in Ar at 600 $^\circ$C for 20 hours in RTP \\
            \hline
            600-20h-B & Anneal in Ar at 600 $^\circ$C for 20 hours in RTP\\
            \hline
            600-16h & Anneal in Ar at 600 $^\circ$C for 16 hours in RTP\\
            \hline
            950-10m-A & Anneal in Ar at 950 $^\circ$C for 10 minutes in RTP\\
            \hline
            950-10m-B & Anneal in Ar at 950 $^\circ$C for 10 minutes in RTP \\
            \hline
            500-600-27h-A  & Anneal at 340 $^\circ$C on hot plate and holding for 10 mins \\
            500-600-27h-A$^1$  & Additional Anneal in Ar at 500 $^\circ$C for 1 hour in RTP  \\
            500-600-27h-A$^2$  &  Additional Anneal in Ar at 600 $^\circ$C for 30 minutes in RTP  \\
            500-600-27h-A$^3$  & Additional Anneal in Ar and H$_2$ at 600 $^\circ$C for 5 hours in RTP \\
            500-600-27h-A$^4$ &  Additional Anneal in Ar at 600 $^\circ$C for 20.5 hours in RTP  \\
            \hline
            500-600-27h-B &  Oxidative Anneal at 340 $^\circ$C holding for 10 mins \\  
            \hline
            \multicolumn{2}{c}{ } \\
            \multicolumn{2}{l}{*All samples Oxidative anneal at 340 $^\circ$C on hot plate and holding for 10 mins}\\        
 
\end{tabular}
\end{table*}

Figure \ref{fig4}(a) shows conductivity of the bridge section as a function of increasing processing time at 600 $^\circ$C and of a single anneal at 950 $^\circ$C in argon (Ar) in RTP. It shows comparison between two separate samples processed up to 16 hours and 20 hours at 600 $^\circ$C as well as between two structures (600-20h-A and 600-20h-B) of same 3D printed dimensions on the same chip. The samples show conductivity of less than 1 S/m if the annealing time is less than 5 hours. However, the conductivity keeps on increasing as the processing time at 600 $^\circ$C rises for both samples. Thus, they can reproducibly show conductivity of up to 20 S/m after 16 hours of anneal at 600 $^\circ$C in Ar. However, the maximum conductivity achieved in this long annealing process is still almost 2 orders of magnitude lower than that of a sample annealed at 950 $^\circ$C for 10 minutes showing that a longer annealing process at 600 $^\circ$C cannot completely pyrolyze the polymer, thereby having post-anneal lower conductivity.

Figure \ref{fig4}(b) shows the measured conductivity of several samples as a function of elapsed days after their processing at 600 $^\circ$C. This shows that the conductivity achieved by longer anneal at 600 $^\circ$C is not constant, rather it decays over time, and it can decay by an order of magnitude in about 40 days. The hypothesis is that by longer annealing, the precursor can reach a metastable state of higher electrical conductivity. By metastability, we are referring to an intermediate energy state within a dynamical system, other than the system’s actual stable state. The experimental conductivity of these polymer structures display such characteristics. Due to prolonged thermal anneal, the sample achieves a higher conductivity, which is a metastable state for the partially carbonized polymer. However, upon removal of thermal energy, the sample’s conductivity reaches a lower value, indicating its stable state. However, as the material stabilizes over time, it can reduce to a lower electrically conductive state. 

\begin{figure}[t]
    \centering
    \includegraphics[width=3in]{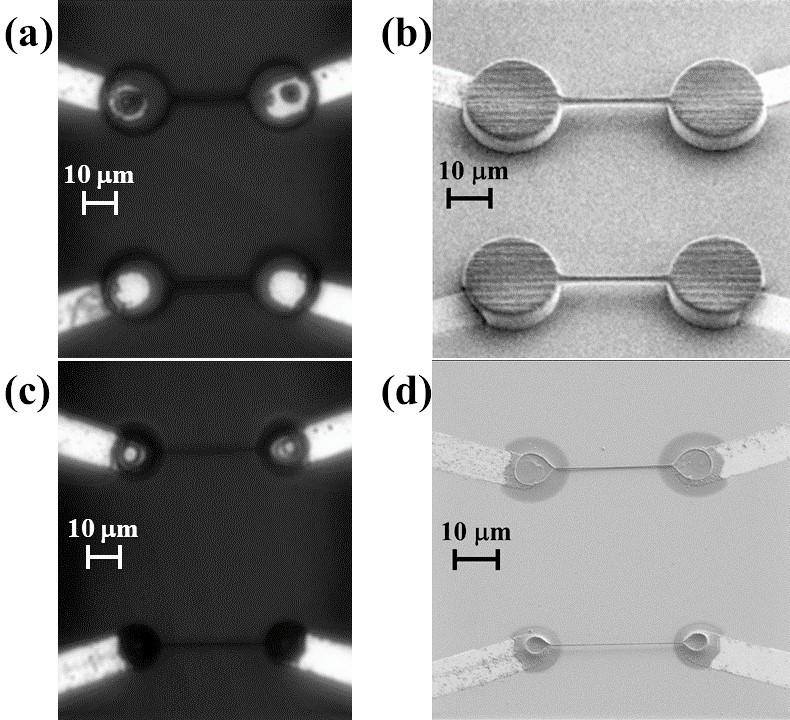}
    \caption{Optical microscopy and SEM images of structures 600-20h-A and 600-20h-B at different steps of processing. (a) Optical microscopy and (b) SEM image of as printed bridge-on-pillars structures standing on metal leads over a silica substrate. (c) Optical microscopy images of structures on silica substrate with metal electrodes after oxidative anneal at 340 $^\circ$C on hot plate for 10 minutes in presence of air. (d) SEM image of structures on silica substrate with metal electrodes at the end of annealing at 600 $^\circ$C in RTP. (Background has been lightened to increase visibility)}
    \label{fig5}
\end{figure}

It is important to confirm the structural stability of the structures and connectivity with the electrodes as the samples are processed at high temperatures and longer times. Figure \ref{fig5} shows optical microscopy and SEM images after various stages of processing of sample 600-20h-A and 600-20h-B. Figure \ref{fig5}(a) shows optical microscopy images of the samples after 3D printing. It shows the structures are optically transparent after 3D printing. Figure \ref{fig5}(b) shows SEM images of the 3D printed structures.  

Figure \ref{fig5}(c) shows optical microscopy images of the structures after oxidative anneal at 340 $^\circ$C for 10 minutes in air. It shows that the structures lose their transparency as a result of oxidation of polymers. Also, significant shrinkage is observed compared to their 3D printed footprint. 

Figure \ref{fig5}(d) shows the SEM image of the bridge-on-pillars structures at the end of their annealing at 600 $^\circ$C for 20 hours in RTP. One important observation here is that after melting and shrinking during annealing, pillars can overlap with metals maintaining uniform contacts and bridge can hang in between the pillars without collapsing. Also, the height of the annealed pillars is dependent on the height of the 3D printed pillars.

\begin{figure}[t]
    \centering
    \includegraphics[width=3in]{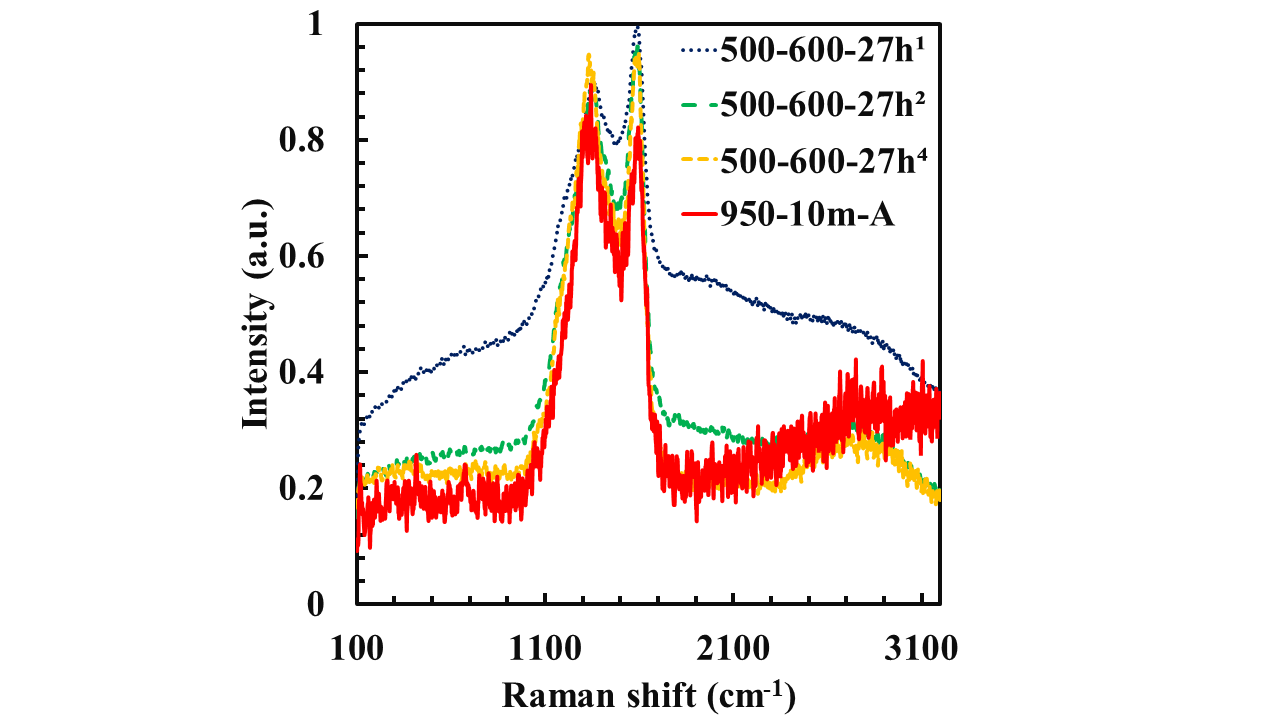}
    \caption{Comparison among Raman spectra at the end of various processing steps of sample 500-600-27h (500-600-27h$^1$, 500-600-27h$^2$, and 500-600-27h$^4$) and at the end of annealing in 950 °C of sample 950-10m-A.}
    \label{fig6}
\end{figure}

Raman spectroscopy was observed to learn about the chemical structure and molecular interactions of the precursor at various stages of carbonization of the precursors. Figure \ref{fig6} shows Raman spectroscopy of the sample 500-600-27h after multiple annealing steps at 500 $^\circ$C and 600 $^\circ$C for 1–27 hours and of sample 950-10m-A after 950 $^\circ$C anneal for 10 minutes. It shows that annealing at only 500 $^\circ$C for up to 1 hour exhibits high background fluorescence. This background fluorescence drops with increasing annealing temperature and time, and after an additional anneal of 2.5 hours at 600 $^\circ$C, the spectrum is quite similar for both 500-600-27h4 and 950-10m-A.

The D-band, also known as the defect band, and the G band are visible at 1345 $cm^{-1}$ and 1595 $cm^{-1}$, respectively. Both D-band and G-band are associated with in-plane vibrations of hexagonal graphitic layers. The D-band is also caused by structural imperfections. The width and relative intensities of these two bands are functions of processing temperatures.

A broad G'-band (first overtone of D-band) occurs at $\approx 2750$ $cm^{-1}$ in 950-10m-A which is absent in sample 500-600-27h (annealed for a relatively shorter time and lower temperature), and weak in longer annealed 500-600-27h \cite{shimada2005origin,vidano1978new}. The Raman spectra does not show any distinguishable feature in the sample annealed at 600 $^\circ$C for 27 hours compared to the sample annealed at 950 $^\circ$C.

\begin{figure}[t]
    \centering
    \includegraphics[width=3in]{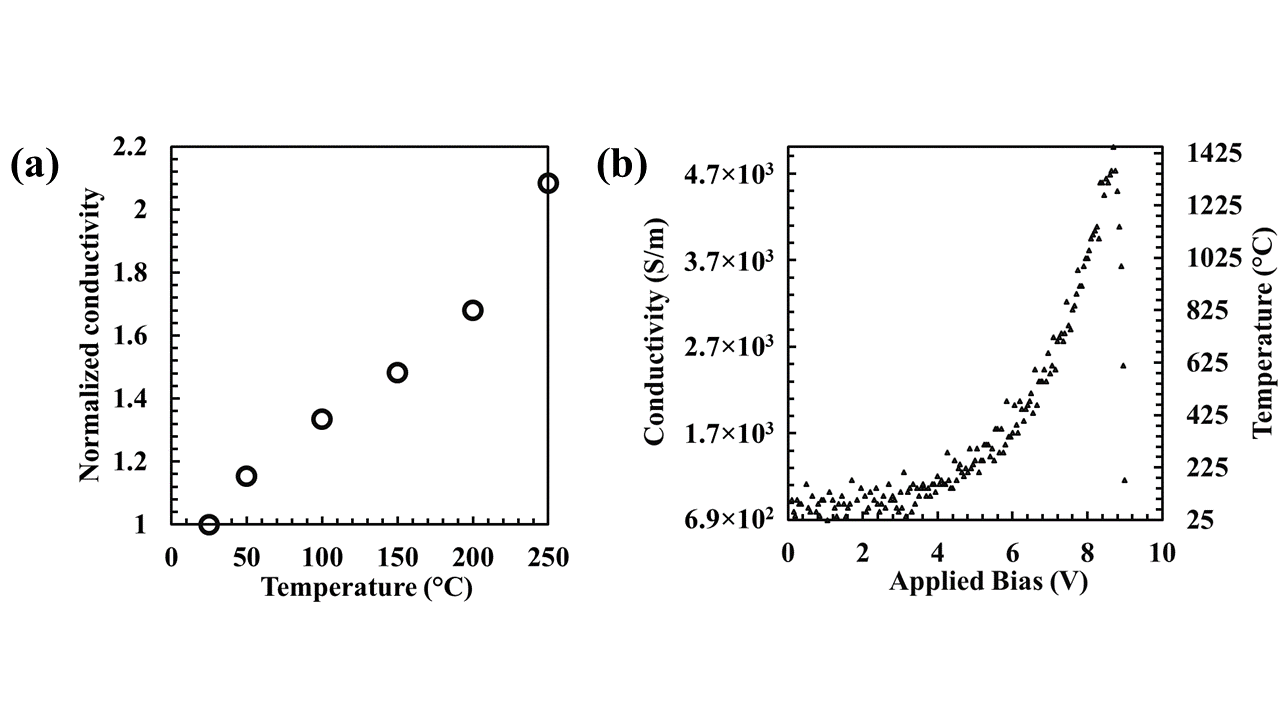}
    \caption{(a) Change in normalized conductivity as a function of ambient temperature (25.2 to 250 °C) of sample 950-10m-A. (b) Conductivity (primary y-axis) and estimated average temperature (secondary y-axis) of the bridge section of sample 950-10m-A as a function of applied bias voltage.}
    \label{fig7}
\end{figure}

A nonlinear change in conductivity due to Joule heating was observed for the sample 950-10m-A. Figure \ref{fig7}(a) shows ambient temperature dependency of normalized conductivity for up to 250 $^\circ$C of the sample that was annealed at 950 $^\circ$C for 10 minutes. For this experiment, the global temperature of the sample is gradually increased from room temperature (25 $^\circ$C) to 250 $^\circ$C. The plot shows a fairly linear trend, and was used to determine the temperature as a function of conductivity for the subsequent experiments to initiate significant Joule heating in a sample annealed at 950 $^\circ$C. 

For the next step of the experiment, a bias voltage is applied across the bridge-on-pillars structure. Figure \ref{fig7}(b) shows the change in conductivity as a function of applied bias in ambient room temperature for a sample annealed at 950 $^\circ$C for 10 minutes. The starting conductivity of the sample was 687 S/m. After approximately 6 V bias, the conductivity increases non-linearly up to the bias voltage of 8.7 V. Beyond 8.7 V, the conductivity drops sharply, indicating damage within the bridge structure. The right y-axis in figure \ref{fig7}(b) shows the estimated average temperature within the bridge as a function of applied bias voltage. It shows that an estimated temperature of $\approx1450$ $^\circ$C was generated within the bridge at the maximum conductivity. The potential reason for the damage in the structure is attributed to evaporation of carbon at the high local temperature within the bridge. This gives a strong indication that an appropriate high electrical conductivity can potentially generate high local temperature in a sample annealed at 600 $^\circ$C by applying a reasonable and safe voltage.

\begin{figure}[t]
    \centering
    \includegraphics[width=3in]{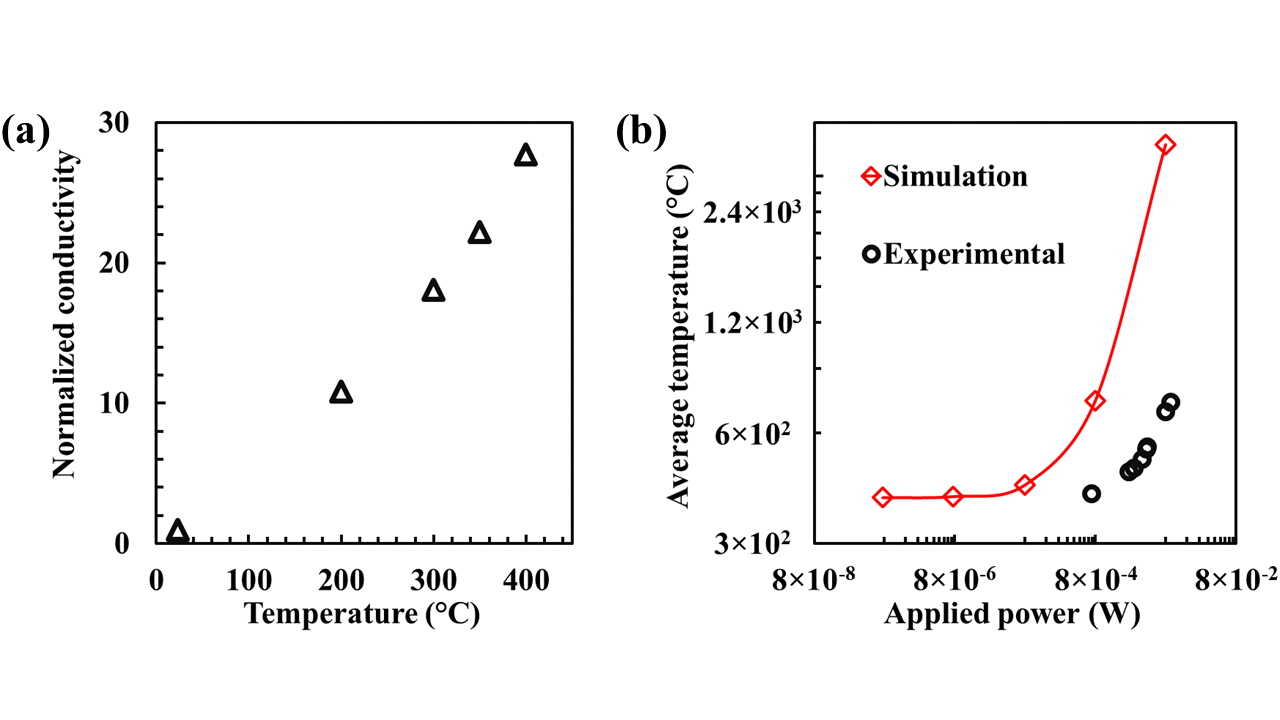}
    \caption{(a) Change in normalized conductivity as a function of ambient temperature (23 to 400 $^\circ$C) of sample 500-600-27h-A. (b) Comparison of simulated and estimated average temperatures of the bridge section of the bridge-on-pillars structure as a function of applied electrical power. For simulation, the bridge-on-pillars structure with pillar 2$\mu$m in height, 4$\mu$m in diameter and bridge 30$\mu$m in length, 0.183$\mu$m in diameter was used having thermal conductivity of 2 W/mK and surface emissivity of 0.8 W/m$^2$. Under an electrical bias of 30 V and in a gaseous environment with a thermal conductivity of 0.05 W/mK and ambient temperature of 400 $^\circ$C, the conductivity of the simulation model was increased to raise electrical power. For experimentally estimating temperature, a 50 k$\Omega$ resistor was kept in series with sample 500-600-27h-A in an ambient temperature of 400 °C while bias voltage was increased up to 31.5 V and current was measured. Data in (a) was used as a calibration curve to estimate average temperature within the bridge.}
    \label{fig8}
\end{figure}

Electrical conductivity of the sample annealed at 600 $^\circ$C (500-600-27h-A) was 9.92 S/m at 23 $^\circ$C. Literature shows that the electrical conductivity of pyrolyzed polymer is analogous to semiconductor material, i.e. the electrical conductivity increases when ambient temperature increases \cite{park2005electrical}. Therefore, the temperature of the sample 500-600-27h-A was increased from 23 $^\circ$C to 400 $^\circ$C and the conductivity measured at 200, 300, 350 and 400 $^\circ$C. This was determined from the linear current-voltage relationship for applied bias. Applied biases were kept under 1 V to make sure that significant Joule heating was not introduced in the system, causing non-linear current-voltage characteristics. Figure \ref{fig8}(a) shows a change in normalized electrical conductivity as a function of increased ambient temperature up to 400 $^\circ$C. It shows a nonlinear increase as temperature is increased, which is consistent with literature \cite{bucker1973preparation}. Figure \ref{fig7} and figure \ref{fig8} also indicate that the conductivity of the 950-10m-A sample is more homogeneous than that of the 500-600-27h-A sample. At a higher temperature, the sample had been pyrolized uniformly. Hence, the difference in normalized conductivity is small. On the other hand, for lower temperature pyrolysis, the difference between conductivity is greater, i.e., as some region of the sample has much higher conductivity than that of other partially pyrolized regions. For best approximation, the last two data points of the curve in figure \ref{fig8} were used to estimate temperature as a function of conductivity for the next experiment.

For this experiment, bias voltages of 28 V and higher were applied across the bridge-on-pillar structure in series with a current limiting resistor (50 k$\Omega$) while the ambient temperature was 400 $^\circ$C and forming gas (2\% H$_2$ + 98\% Argon) was flowing in the micro-hotplate chamber to prevent over-oxidation of the precursor at high temperatures. Figure \ref{fig8}(b) shows the estimated average temperature within the bridge as a function of the applied power and compares this result with finite-element-based COMSOL simulation. As the plot shows, the bridge reaches up to a temperature of 727 $^\circ$C due to current-driven Joule heating. However, it needs to be pointed out here that this estimated temperature is the average-calculated temperature within the bridge. Therefore, this underestimates any higher temperature region within the bridge that is caused by non-uniformity in bridge thickness. For the simulation, a thermal conductivity of 2 W/mK was used based on literature \cite{venkatachalam2017thermal} and a surface emissivity of 0.8 W/$m^2$ was used for the partially pyrolyzed polymer. 

The calibration curve in figure \ref{fig8}(a) was experimentally derived to estimate the temperature from the conductivity for Fig 8(b). This may be a source of  discrepancy between the experimental and simulation curves. Furthermore, the spacing between the substrate and annealed bridge can vary depending on initial pillar height. This spacing determines the volume of the gas that absorbs the heat dissipating from the bridge. Depending on this spacing, the heat reflects from the substrate towards the bridge section. Thus, this spacing can also affect the steady state temperature within the bridge during Joule heating. This process is challenging to control due to current runaway conditions when the bridge goes into nonlinear conductivity region. At these runaway conditions, the bridge is prone to damage. Use of a constant current source can potentially provide better control over the current flow. Instead of starting with a higher conductivity bridge-on-pillars structure, irreversible conductivity can also be achieved by using higher voltage supply. However, that approach may be limited due to not being compatible with CMOS technology.

\begin{figure}[t!]
    \centering
    \includegraphics[width=3in]{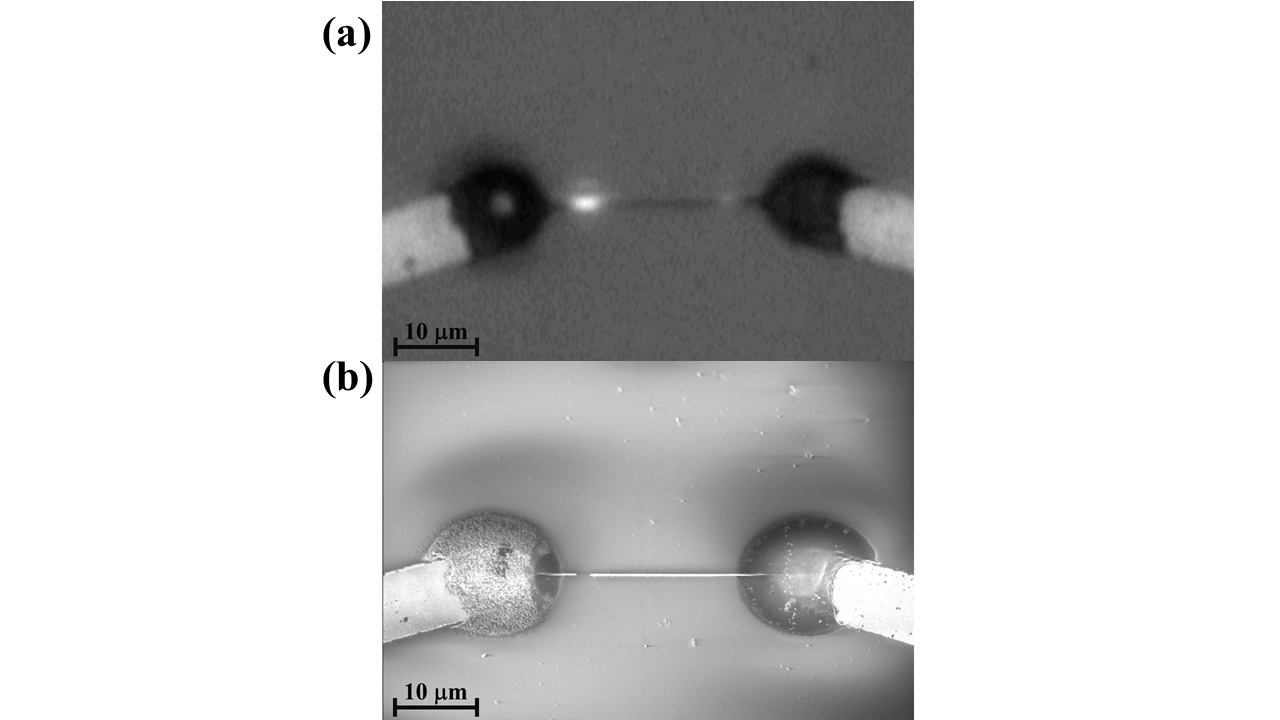}
    \caption{(a) Optical image of sample 500-600-27h (connected in series with a 50 k$\Omega$ resistor, not shown in the image) under a bias voltage of 31.6 V, while the ambient temperature was 400 $^\circ$C. (b) SEM image of sample 500-600-27h after bridge section broke as a result of generating high temperature within the bridge from driving high current of $\approx$300 $\mu$A while the ambient temperature was 400 $^\circ$C.}
    \label{fig9}
\end{figure}

Figure \ref{fig9}(a) shows an optical microscopy image of the sample 500-600-27h for a bias voltage of 31.6 V, series resistance of 50 k$\Omega$, and measured current of 300 $\mu$A. The image was taken in a dark ambience. A bright light spot was observed within the bridge, which shows a significant temperature increase compared to the dark 400 $^\circ$C ambience. This also shows that the wire is not uniform in thickness, thereby generating the most temperature in the narrowest region. However, this bright spot was gone after about 20 minutes and the current dropped to zero, indicating damage/discontinuity in the structure.  

Figure \ref{fig9}(b) shows the SEM of sample 500-600-27h after the current dropped to zero, which clearly shows a breakage within the bridge section. From the SEM images, we can also see that although the surface of the pillars is rough after Joule heating, the active bridge portions are relatively smoother and physically stable. The diameter of annealed carbon nanowires typically range between 200 and 300 nm. 

\section{Conclusion}
Lab-on-CMOS electrochemical sensors are potentially viable solutions for commercialized sensing products. Custom-designed lab-on-CMOS systems offer the advantage of individually addressed transducers or sensing elements, have relatively low cost in fabrication, and high sensitivity. It can also potentially minimize interference from the ambient environment when the transducer is tightly packed with the readout electronics. 

In this work, optimized processing conditions, including ambient temperature, starting electrical conductivity of partially pyrolyzed polymer, and thermal conductivity of ambience, have been explored to convert partially pyrolyzed 3D printed polymer microstructures to completely carbonized microelectrodes using an electrical current. The novelty of the developed processing sequence lies in experimentally demonstrating the feasibility of achieving complete carbonization of partially pyrolyzed polymer using a Joule heating approach.  The proposed two-step anneal process with relatively short duration (from 10 minutes to 2 hours), second anneal at 500-600 $^\circ$C produced partially carbonized structures with conductivity too low to be useful for localized Joule heating. Only some of the longest anneals with durations of 16–27 hours resulted in a conductivity sufficient for subsequent localized Joule heating anneals. We also found that the relatively high conductivity achieved after such prolonged anneals was not stable, but rather decreased substantially with time during sample storage. I-V curves under varying ambient temperature were measured in the range of 20-350 $^\circ$C and analyzed to deduce thermal coefficients of electrical conductivity. This allowed us to use the structure itself as a thermistor and estimate its actual temperature by measuring its resistance. An estimated temperature of 727 $^\circ$C was achieved within the active region of the structure by applying a safe voltage. This approach showed that localized Joule heating can be initiated within this bridge-on-pillars structure by driving current, thus showing feasibility of carbonization without raising the temperature of the entire chip above 600 $^\circ$C.  In summary, the scientific contribution of this work is in successfully demonstrating a processing sequence that reduces annealing temperature of carbonizing electrodes within CMOS tolerable temperatures.

\section{Acknowledgment}
Design, fabrication and characterization of the samples involved in this research was conducted at the Center for Nanophase Materials Sciences, which is a DOE Office of Science User Facility.

\bibliographystyle{plain}
\bibliography{references}

\end{document}